\documentclass[twocolumn,superscriptaddress,longbibliography,
	aps,pra,preprintnumbers,floatfix]{revtex4-2}

\usepackage{graphicx}
\usepackage{dcolumn}
\usepackage{bm}
\usepackage{amsmath}
\usepackage[colorlinks=true, citecolor=red, linkcolor=blue, urlcolor=black]{hyperref}

\usepackage{enumerate}
\usepackage{float}

\def\d{{\textit{d}}}

\usepackage[dvipsnames]{xcolor}
\usepackage[normalem]{ulem}
\usepackage{ifthen}

\newcounter{verbosity}
\begin{document}
\title{Long-range order in the XY model on the honeycomb lattice}

\author{      Jacek Wojtkiewicz}
\affiliation{ \mbox{Institute of Theoretical Physics, Faculty of Physics,
              University of Warsaw, ul. Pasteura 5, PL-02093 Warsaw, Poland} }
\author{      Krzysztof Wohlfeld}
\affiliation{ \mbox{Institute of Theoretical Physics, Faculty of Physics,
              University of Warsaw, ul. Pasteura 5, PL-02093 Warsaw, Poland} }
\author{      Andrzej M. Ole\'s$\,$}
\email[e-mail: ]{a.m.oles@fkf.mpg.de}
\affiliation{Max Planck Institute for Solid State Research,
Heisenbergstrasse 1, D-70569 Stuttgart, Germany}
\affiliation{\mbox{Institute of Theoretical Physics, Jagiellonian University,
Prof. Stanis\l{}awa \L{}ojasiewicza 11, PL-30348 Krak\'{o}w, Poland}}

\date{\today}

\begin{abstract}
Using the reflection positivity method we provide the rigorous proof of
the existence of long range magnetic order for the XY model on the
honeycomb lattice for large spins $S\ge 2$. This stays in contrast with
the result obtained using the {\it same} method but on the square
lattice---which gives a stable long-range order for spins $S\ge 1$.
We suggest that the difference between these two cases stems from the
enhanced quantum spin fluctuations on the honeycomb lattice. Using
linear spin-wave theory we show that the enhanced fluctuations are due
to the overall much higher kinetic energy of the spin waves on the
honeycomb lattice (with Dirac points) than on the square lattice
(with good nesting properties).
\end{abstract}

\maketitle

\section{Introduction}

One of the main questions concerning the nature of a quantum spin model
is whether long-range order (LRO) of any kind could stabilize in a
certain range of parameters. Models with finite magnetic anisotropies
have discrete symmetries and order much easier, which agrees with our
intuition. An occurrence of phase transition in the two-dimensional (2D)
Ising model, established on the basis of exact solution of interacting
spins $S$ at a square lattice is the most prominent example. The LRO
sets in the Ising model at positive temperature $T<T_c$, where $T_c$ is
the critical temperature, though the order parameter is reduced by
thermal fluctuations~\cite{Lars}. The spin order parameter becomes
maximal in the ground state, i.e., $\langle S^z\rangle\equiv S$ at $T=0$.
On the other hand, the one-dimensional (1D) Ising model orders
{\it only} at $T=0$.

Models with continuous symmetries, such as the Heisenberg
SU(2)-symmetric or the XY U(1)-symmetric one, are quite different.
In this case the Mermin-Wagner theorem prevents the LRO at positive
temperatures in `all' low dimensions, i.e., not only in a 1D but also
in a two-dimensional (2D) model~\cite{Mer66, Auerbach1994, Bee19}.
This is due to the proliferation of the Goldstone modes, i.e., gapless
spin waves, which enhances the thermal fluctuations and kill the LRO.
Moreover, the $T=0$ version of the above theorem, the Coleman theorem
\cite{Bee19}, states that at $T=0$ the ground state of such models
`typically'~\footnote{It means in magnets with type-A Goldstone
modes~\cite{Bee19}. Note that below we consider solely magnets with such
Goldstone modes, i.e., we exclude those with type-B Goldstone models
(as best exemplified by the Heisenberg ferromagnet).}
does not carry LRO in one dimension, for the order parameter is
completely destroyed by quantum fluctuations triggered by the spin
waves. This happens for instance in the 1D antiferromagnetic (AFM)
Heisenberg model~\cite{Bet31, Auerbach1994}.

This shows that the 2D spin models with continuous symmetries are
special, since they never order at positive temperature but may order at $T=0$.
Note that the Coleman theorem merely {\it allows} for the onset of the LRO
but does {\it not} guarantee it. This is probably best exemplified
by the search for the spin liquid ground state of Heisenberg models on
2D frustrated lattices~\cite{Leon,Saba}. But an
intriguing situation arises already in the 2D AFM Heisenberg model
on the square lattice. Here the linear spin-wave theory (LSWT)
expansion suggests that at least for a `large-enough' size of spin $S$
the quantum corrections to the order parameter ($\Delta m$) should be small enough to allow for LRO. In fact, several numerical and analytical methods suggest that the LRO exists (already) for spin $S=1/2$~\cite{Sin89, Ham92, Whi07, San10, Kad22}. However, it is both an amusing and irritating circumstance
that its existence has not been rigorously confirmed:
the proof shows that the LRO can be stable only for spins
$S\ge 1$~\cite{KLS_JSP}. This fragility of the long-range spin order is
further strengthened by the research on the low-lying excited states which
suggests that these may partially be better described in terms of
excitations from a spin liquid state (spinons)~\cite{Syl02, Pia15, Bec18} than from the symmetry-broken state with LRO (spin waves or magnons)~\cite{Pow15, Pow18, Ver18}.

Experimentally,
the 2D spin models with continuous symmetries can basically be realised in
the van der Waals crystals \mbox{\cite{Gon17,Bur18,Gon19}.} These materials
are a subject of intensive research and one of the main questions is whether
the thermal fluctuations kill the magnetic order. In principle this should
not be the case, for the assumptions of the Mermin-Wagner theorem are never
strictly fulfilled in the materials that are solely {\it quasi}-2D and
always have finite magnetic anisotropies. Yet, the applicability of the
Mermin-Wagner theorem as well as the potential onset of the large
thermal corrections to the order parameter are a subject of intensive
discussion~\mbox{\cite{Gon17,Bur18,Gon19}.} It is thus natural to
investigate what the role played by the quantum fluctuations is in the
onset of the LRO in the ground state of such a `van der Waals' spin
model which is approximately 2D and has continuous symmetry. Here the
salient feature is that the van der Waals materials support the
honeycomb~\mbox{\cite{
Gon17,Bur18,Gon19, Joy92,Kimnc,Kim19,Kim20,Coa21,Waals}},
rather than the square, lattice. Although honeycomb lattice is bipartite,
just as the square lattice is, it is not obvious {\it how the LRO spin
order would survive in the ground state of a spin model with continuous
symmetry on the honeycomb lattice at $T=0$}---which is the main question
of this work.

Indeed, the spin LRO is likely to be less stable
on the honeycomb than for the square
lattice as there are three outgoing exchange bonds from each site and this
may amplify the effects of quantum fluctuations which could destroy the
ordered state. (Notably, for the 1D Heisenberg model there are just two exchange bonds outgoing from each site and the LRO is then destroyed,
as just discussed above.) In fact,
for the Heisenberg model on the honeycomb lattice there is
a proof of the existence of LRO for $S\ge 3/2$ \cite{AKLT88} whereas for the
square lattice case the {\it analogous} proof applies already to spins $S\ge 1$ \cite{KLS_JSP}, see above.
Another argument in favor of diminished stability of the ordered AFM state
on the honeycomb lattice is provided by the LSWT: Quantum corrections
to the order parameter $\Delta m$ for the Heisenberg model are substantially larger on the honeycomb lattice than on the square lattice, \mbox{(i.e.,
$\Delta m \simeq 0.28$ versus $\Delta m \simeq 0.197$ \cite{Mattis}).}

In this work we shall investigate the existence of LRO on
the honeycomb lattice in the spin model and discuss why the order may be
less stable than on the square lattice.
To this end we choose to work with the XY model. While this model
has a lower symmetry than the Heisenberg model
\cite{Gom87} and one expects that it orders easier,
there is no proof of the existence of the LRO using the same method
as the one used for its square lattice counterpart and yielding LRO
for $S\ge 1$~\cite{Kubo} either~\footnote{In fact, the LRO has been
proven for $S \ge 1/2$ for the square lattice~\cite{KLS}---but
using a distinct method than the one here, see discussion in Section~\ref{sec:rp}.}.
It is thus of crucial importance to verify whether the LRO is stable in this model for the same {\it or} for the higher spin $S$ value than on the square
lattice. Moreover, the LSWT corrections calculated to the order parameter
of the XY model for the honeycomb lattice are also larger than for the
square lattice (i.e., $\Delta m \simeq 0.08$ versus $\Delta m \simeq 0.06$ \cite{Gom87,Wei91}), making it an ideal case to compare the stability
of the LRO on these two distinct 2D lattices.

The paper is organized as follows. We shall start from the reflection
positivity (RP) method applied to the XY spin model on the honeycomb
lattice in Sec. \ref{sec:rp}. Next we use the LSWT method to intuitively
understand why the LRO on the honeycomb lattice is
less stable than on the square lattice, see Sec. \ref{sec:lswt}.
The summary and conclusions are presented in Sec.~\ref{sec:summa}.

\section{Results:\\
Reflection positivity method}
\label{sec:rp}

In this chapter, we rigorously prove that N\'eel order exists in the
ground state of the XY model on the honeycomb lattice with AFM
interactions, for a large enough value of spins. The proof is
based on the reflection positivity (RP) technique.
Note that, while we consider below solely the AFM case,
the proof is valid for the ferromagnetic XY model as well---since
these models are in fact isomorphic, unlike in the Heisenberg case.

For quantum spin systems, the RP technique originated in a seminal paper
\cite{DLS}. This paper treated spin systems in three dimensions at
positive temperature. In two dimensions, there is no LRO at positive
temperature in systems with continuous symmetry group due to
Mermin-Wagner theorem \cite{Mer66}, but there remains non-trivial
question of the ground-state ordering. In the paper \cite{NevesPerez}
authors have shown that technique of Ref. \cite{DLS} can be adapted to
prove the existence of N\'eel order in the ground state of an AFM
Heisenberg model on the square lattice provided $S\geq 3\slash{}2$.
Later on, it was noticed that the authors of Ref. \cite{NevesPerez}
made a numerical error and in fact LRO exists for all spins $S\geq 1$
\cite{KLS_JSP}. Similarly, the LRO in the Heisenberg model on the honeycomb lattice was proven for spins $S\ge 3/2$ \cite{AKLT88}.

For reasons presented in the previous Section, it would be desirable to
settle the question of ground-state ordering for XY model on the honeycomb
lattice. This problem has been answered in positive manner for the
square lattice:
whereas using an {\it analogous} method as presented below it was shown
that the LRO exists in the ground state for spins $S\ge 1$ \cite{Kubo},
a distinct calculation showed that
the ground state is ordered for arbitrary value of spin~\cite{KLS}.
However, we are not aware of such result for honeycomb lattice, and this
opportunity encouraged us to undertake attempts to prove this. Below we
supply such a proof. The calculation is
based on an adaptation of the AKLT technique \cite{AKLT88},
with heavy use of results given in Ref. \cite{DLS}, as well as in
Ref. \cite{NevesPerez}, so we do not include here all the details.

We write the Hamiltonian as
\begin{equation}
{\cal H}_\Lambda= \sum_{\langle {\bf m},{\bf n}\rangle}
h^{\rm YZ}_{\langle {\bf m}{\bf n}\rangle}\,.
\label{YZ}
\end{equation}
where ${\bf m}$ and ${\bf n}$ are two connected sites,
$\langle {\bf m},{\bf n} \rangle$ is a bond
between nearest neighbors, and the summation is performed over
such bonds on the honeycomb lattice, and we define
\begin{equation}
h^{\rm YZ}_{\langle {\bf m}{\bf n}\rangle}=
S^2_{\bf m} S^2_{\bf n} + S^3_{\bf m} S^3_{\bf n}\,.
\label{HamXY2spins}
\end{equation}
Note that we take for convenience exchange couplings between the
components 2 and 3
(or $\{\rm Y,\rm Z\}$, respectively) instead of conventional 1 and 2
(or $\{\rm X,\rm Y\}$). Both models are of course physically equivalent.
Our choice is dictated by an easier comparison with results for an
isotropic Heisenberg model. In particular, we define as the order
parameter the average of the 3$^{rd}$ (or $Z^{th}$) component, in
analogy to \cite{AKLT88}.

The first step of calculation is to compute eigenvalues and
eigenvectors of the Laplacian $-\Delta$ on the honeycomb lattice.
It is defined as
\begin{equation}
(-\Delta\psi)({\bf m}) = 3\psi({\bf m})
-\sum_{{\bf n}:||{\bf m}-{\bf n}||=1}\psi({\bf n}),
\label{Lapl}
\end{equation}
where $||{\bf m}-{\bf n}||=1$ means that $\{{\bf m},{\bf n}\}$ are
nearest neighbor sites.
The honeycomb lattice is periodic with period 2, so eigenvectors
cannot be found by ordinary Fourier transform. Instead, observe that
the honeycomb lattice is bipartite, i.e., it can be represented as a
disjoint union of two sublattices, with any nearest neighbors
belonging to different sublattices. We will refer to these sublattices
as `even' and `odd' ones; they are yet strictly periodic. One performs
two Fourier transforms associated with these sublattices.

Let ${\boldsymbol\delta}_i$, for $i=1,2,3$, are unit lattice vectors such
that for every even site ${\bf m}$, the nearest neighbors are given by
\mbox{${\bf m}+{\boldsymbol\delta}_i$}. The nearest neighbors of
an odd site ${\bf n}$ are then ${\bf n}-{\boldsymbol\delta}_i$.
We take explicitly: ${\boldsymbol\delta}_1=(0,-1)$;
${\boldsymbol\delta}_2=(\frac{\sqrt{3}}{2},\frac{1}{2})$;
\mbox{${\boldsymbol\delta}_3=(-\frac{\sqrt{3}}{2},\frac{1}{2})$.}
For a finite lattice $\Lambda$ with periodic boundary conditions the
eigenvalues of Laplacian are grouped in two `$\pm$ bands':
\begin{equation}
    E^{\pm}_{\bf k} = 3\pm|\epsilon({\bf k})|,\quad where \quad
\epsilon({\bf k})=\sum_{j=1}^3\exp\left(i{\bf k}\cdot {\boldsymbol\delta}_j\right).
\label{Epm}
\end{equation}
In an explicit manner:
\begin{align} \label{eq:structfact1}
\epsilon({\bf k}) = \exp (-i\,k_2) + 2 \exp\left( i\,\frac{k_2}{2}\right) \cos\left (\frac{\sqrt{3} k_1} {2} \right),
\end{align}
and therefore
\begin{equation}
|\epsilon({\bf k})|=\sqrt{1+4\cos\frac{\sqrt{3}k_1}{2}
\left(\cos\frac{\sqrt{3}k_1}{2} + \cos\frac{3 k_2}{2}\right)}.
\label{epsk}
\end{equation}

Here, the momentum ${\bf k}\equiv(k_1,k_2)$ takes values in the Brillouin
zone (BZ) for one of the two sublattices. Corresponding eigenvectors
$h^\pm_{\bf k}({\bf m})$ are:
\begin{eqnarray}
h^+_{\bf k}({\bf m})&=&{\rm{sgn}}({\bf m})\frac{1}{\sqrt{|\Lambda|}}
\exp\!\left[i{\bf k}\!\cdot\!{\bf m}+{\rm{sgn}}({\bf m})
\frac{\phi({\bf k})}{2} \right]\!,\\
h^-_{\bf k}({\bf m})&=&\frac{1}{\sqrt{|\Lambda|}}
\exp\!\left[i{\bf k}\cdot{\bf m} +{\rm{sgn}}({\bf m})\frac{\phi({\bf k})}{2} \right],
\end{eqnarray}
where the phase $\phi({\bf k})$ is determined from
\[
\epsilon({\bf k})=|\epsilon({\bf k})|\exp(i\phi({\bf k})).
\]
Following \cite{DLS}, and in particular Lemma 6.1 therein, one proves the
{\em Gaussian domination} inequality common for both Heisenberg and YZ models:
\begin{eqnarray}
&&\left(\sum_{\bf m}{\rm{sgn}({\bf m})S^3_{\bf m}\overline{
(-\Delta f)({\bf m})}\times
\sum_{\bf n}\rm{sgn}}({\bf n})S^3_{\bf n}{(-\Delta f)({\bf n})}
\right)\nonumber\\
&&\leq
\beta^{-1}\sum_{\bf m}\overline{f({\bf m})}(-\Delta f)({\bf m})\,,
\label{GaussDom}
\end{eqnarray}
where $(\cdot,\cdot)$ is Duhamel two-point function \cite{DLS}, the bar denotes
the complex conjugation, $f({\bf m})$ is arbitrary function on the lattice,
and $\beta$ is inverse temperature (below we shall take the limit
$\beta\to\infty$).

Define:
\begin{equation}
S^\pm_{\bf k}= \sum_{\bf m} h^\pm_k({\bf m})S^3_{\bf m}.
\label{Spmk}
\end{equation}
(Note that $S^\pm_{\bf k}$ are {\it not} related to the spin raising and
lowering operators.) The sum in Eq. (\ref{Spmk}) includes only third
components of spin operator, and $\pm$ is a band index. Choosing now
eigenvectors $h^\pm_{\bf k}$ as a function $f$ in Eq. (\ref{GaussDom}),
we get
\begin{equation}
(\overline{S^\pm_{\bf k}},S^\pm_{\bf k})\leq\frac{1}{\beta E^\mp_{\bf k}}\,.
\end{equation}
So far, the calculations for YZ model (\ref{YZ}) are very similar to those for
the Heisenberg model \cite{AKLT88}. Substantial difference appears when one
calculates the average of the {\em double commutator}
$\langle [ {S^\pm_{\bf k}}, [H, \overline{S^\pm_{\bf k}}]]\rangle$.
The double commutator is calculated in a standard manner:
\begin{eqnarray}
  & &   [ {S^\pm_{\bf k}}, [H, \overline{S^\pm_{\bf k}}]]
 \nonumber\\
 &=&\frac{1}{|\Lambda|} \sum_{\bf m}\sum_{{\bf n}:||{\bf m}-{\bf n}||=1}
 \left(
 e^{i{\bf k}\cdot({\bf m}-{\bf n})}S^1_{\bf m} S^1_{\bf n}\!
 - S^2_{\bf m} S^2_{\bf n} \right).
\label{eq:dc}
\end{eqnarray}
The average of $\langle S^2_{\bf m}S^2_{\bf n}\rangle$ is expressed
easily by the ground-state energy per site $E_{gs}$:
\begin{equation}
\left\langle S^2_{\bf m}S^2_{\bf n}\right\rangle=
\frac{1}{2}\left\langle h^{\rm YZ}_{\langle{\bf m}{\bf n}\rangle}\right\rangle=
\frac{1}{2}\frac{|\Lambda|}{\cal{N}} E_{gs}\,,
\end{equation}
where $\cal{N}$ is the number of bonds of the lattice. The average of
$\langle S^1_{\bf m} S^1_{\bf n}\rangle$ can be estimated as in Refs.
\cite{KLS_JSP,Kubo}:
\begin{equation}
|\langle S^1_{\bf m} S^1_{\bf n}\rangle| \leq
\left|\left\langle S^2_{\bf m} S^2_{\bf n}\right\rangle\right|\,.
\end{equation}
Putting all together, the average of double commutator can be
estimated as
\begin{equation}
 \langle [ {S^\pm_{\bf k}}, [H, \overline{S^\pm_{\bf k}}]]
\rangle
\leq
\frac13\,\tilde{E}_{\bf k}\,|E_0|,
\end{equation}
where
\begin{equation}
    \tilde{E}_{\bf k}\!= 3+
 \sqrt{1+4\left|\cos\frac{\sqrt{3}k_1}{2}\right|
\left(\left|\cos\frac{\sqrt{3}k_1}{2}\right|
+ \left|\cos\frac{3 k_2}{2}\right|\right)},
\end{equation}
and $E_0$ is an {\em lower bound} on ground state energy. We
have also used $\cos x \leq |\cos x|$ and ${\cal{N}}/|\Lambda|=3/2$
for the honeycomb lattice.

\begin{figure*}[t!]
    \centering
    (a) \includegraphics[width=0.62\columnwidth]{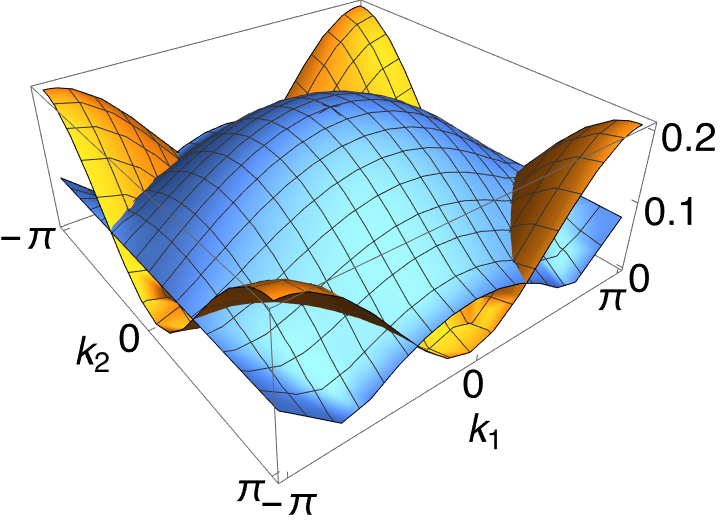}
    (b) \includegraphics[width=0.62\columnwidth]{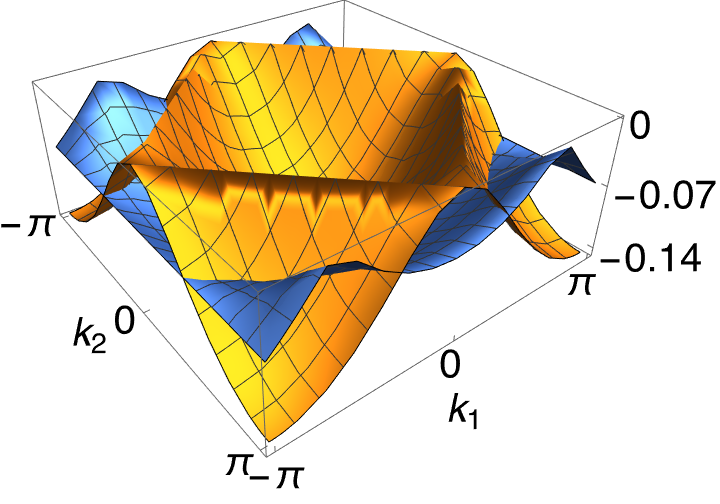}
    (c) \includegraphics[width=0.62\columnwidth]{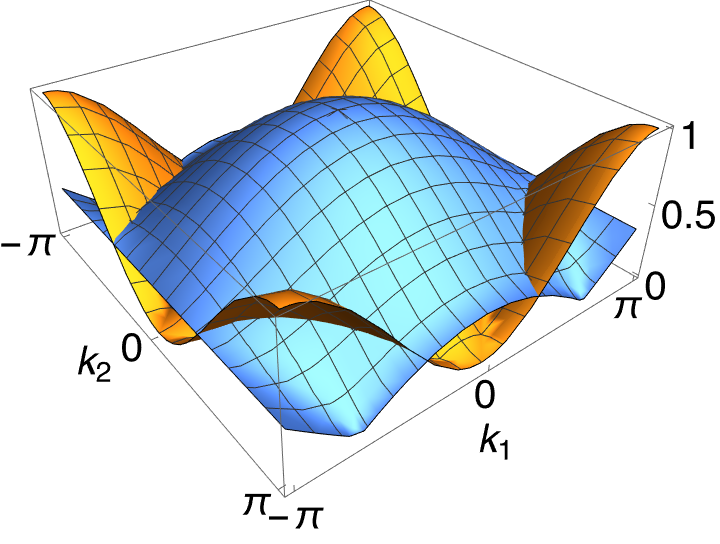}
    \\
    (d) \includegraphics[width=0.62\columnwidth]{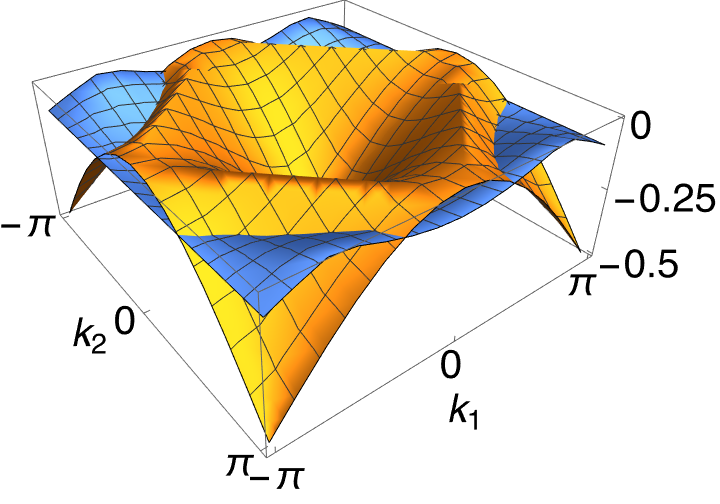}
    (e) \includegraphics[width=0.62\columnwidth]{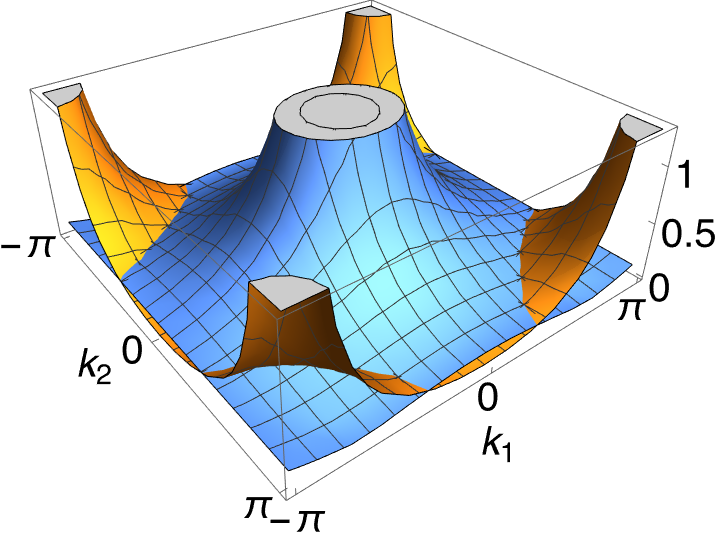}
    (f) \includegraphics[width=0.62\columnwidth]{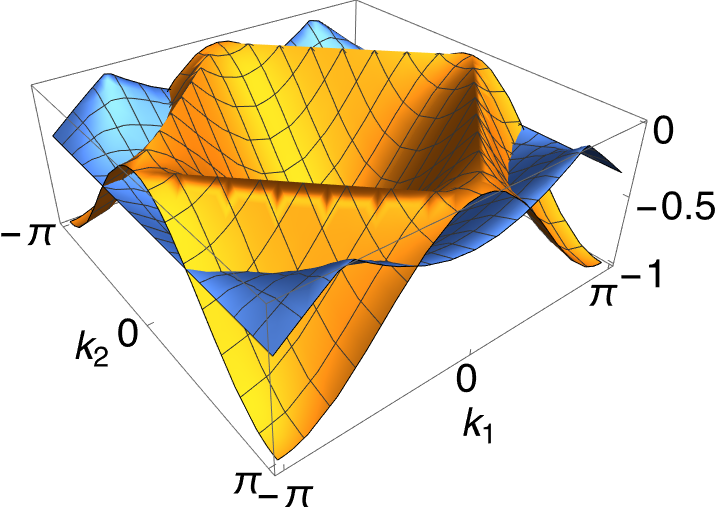}
\vskip .5cm    
\caption{Dependence of the four contributions to the order parameter
renormalization $\Delta m$ \eqref{eq:deltamlswt} on the momentum components
$\{k_1,k_2\}$ in the LSWT approach:
(a) $C^+_1({\bf k})$, (b) $C^+_{-1}({\bf k})$,
(d) $C^-_1({\bf k})$, (e) $C^-_{-1}({\bf k})$.
For comparison, the relevant moduli of the structure factors are also
shown in:
(c) $|\gamma({\bf k})|$ and (f) $-|\gamma({\bf k})|$
(see text for further details). The blue (yellow) color indicates for
the results found for the honeycomb (square) lattice, respectively.}
    \label{fig:1}
\end{figure*}

A crude approximation for $E_0$ can be obtained with aid of
inequalities, expressing the free energy of the YZ model by the
free energy of the Ising model, in a similar manner as in \cite{Simon},
see page 57,
\[
f^{Is}(\beta)\geq f^{\rm YZ}(\beta)\geq f^{Is}(2\beta).
 \]
Passing to the limit $\beta\to\infty$, one obtains:
\begin{equation}
 E^{\rm YZ}_{gs}\geq E^{Is}_{gs}=-2S^2\cal{N},
\end{equation}
so for the honeycomb lattice one finds the lower bound $E_0$ for
the ground state energy per site:
\begin{equation}
    E_0=-3S^2.
\end{equation}
Following the arguments of Refs. \cite{DLS} and \cite{AKLT88}, i.e.,
the Gaussian domination inequality (\ref{GaussDom}), and the Falk-Bruch
inequality, the usual sum rule and
\mbox{$\langle S^3_{\bf m}S^3_{\bf m}\rangle=
\langle\mathbf{S}_{\bf m}\!\cdot\!
\mathbf{S}_{\bf m}\rangle\slash{}3=S(S+1)\slash{}3$}
imply that there is N\'eel order in the ground state if
\begin{equation}
    \frac{2S(S+1)}{3}>S(I_+ + I_-)\,,
\label{MainIneq}
\end{equation}
where
\begin{equation}
I_\pm = \lim_{|\Lambda|\to\infty}\,\sum_{\bf k}\,\sqrt{
\frac{\tilde{E}_{\bf k}}{E^\pm_{\bf k}}
}
=
\frac{1}{2|BZ|}\int_{\rm{BZ}} \d^2 {\bf k}\,
\sqrt{\frac{\tilde{E}_{\bf k}}{E^\pm_{\bf k}}
}\,.
\end{equation}
Here BZ stands for the Brillouin zone for one of two sublattices and
$|BZ|$ is its area. The integrand of $I_+$ is a regular function,
whereas integrand of $I_-$ possess singularities, but they are
integrable. One finds a numerical value of $I_++I_-=1.777$, which
implies that the condition (\ref{MainIneq}) is fulfilled for $S=2$,
or larger. In the other words, we have proved that AFM XY
{\em model on honeycomb lattice possesses N\'eel order in the ground
state for} $S\ge 2$.
We suggest that future research should allow to improve an estimation
for average of double commutator \eqref{eq:dc},
as well as for the upper bound of the ground state energy.

\section{Discussion:\\Linear spin wave theory insight}
\label{sec:lswt}

In this section we would like to provide better understanding why,
as the RP result suggests, the LRO seems to be `softer' on the
honeycomb than on the square lattice
(since the spin for which the LRO exists is proved to be higher for the honeycomb than for the square lattice).
To this end we analyse in detail the well-known LSWT result which gives the quantum correction to the order parameter on the square (honeycomb) lattice $\Delta m =0.06$ ($\Delta m = 0.08$), respectively~\cite{Wei91}. Thus,
we follow the calculations of Ref.~\cite{Wei91} and write down the expression
for the LSWT quantum corrections to the order parameter by performing the
Holstein-Primakoff expansion around the N\'eel state to bosonic operators for
each spin at site ${\bf m}$ of the $\uparrow$-spin sublattice \cite{Mattis}:
\begin{equation}
S_{\bf m}^-\simeq a^{\dagger}_{\bf m}, \quad
S_{\bf m}^+ \simeq a^{}_{\bf m}, \quad
S_{\bf m}^z=S-a_{\bf m}^{\dagger}a_{\bf m}^{},
\end{equation}
and similarly for $\downarrow$-spin sublattice.
Keeping only the linear terms in the bosonic operators
and performing successive Fourier and Bogoliubov transformations,
one finds then the quantum corrections to the order parameter~\cite{Wei91},
\mbox{$\langle S^z_{\bf m}\rangle= S-\Delta m$,} where
\begin{align}
\label{eq:deltamlswt}
\Delta m = \frac14\,\left|\,C^+_1 + C^-_1 +C^+_{-1}  + C^+_{-1}\,\right|\,.
\end{align}
Here
\begin{align}
C^\pm_1 = \frac{z}{2\mathcal{N}} \sum_{\bf k}
C^\pm_1 ({\bf k}), \quad
C^\pm_{-1} = \frac{z}{2\mathcal{N}} \sum_{\bf k}
C^\pm_{-1} ({\bf k}),
\end{align}
and
\begin{align}\label{eq:c1}
C^\pm_1 ({\bf k}) &= \sqrt{1 \pm |\gamma({\bf k})|} - 1, \\
\label{eq:c-1}
C^\pm_{-1} ({\bf k}) &= \frac{1}{\sqrt{1 \pm |\gamma({\bf k})| }} - 1.
\end{align}
A constant $z$ stands for the coordination number of the lattice
($z=3$ for the honeycomb and $z=4$ for the square) and $\gamma(k)$
is the structure factor that for the honeycomb lattice is given by
\begin{align}\label{eq:structfact2}
\gamma({\bf k}) = \frac13 \left[ \exp\left(-i\,\frac{2k_2}{3}\right)
+ 2 \exp\left( i\,\frac{k_2}{3}\right) \cos \left(\frac{k_1}{2}\right)\right],
\end{align}
and for the square lattice by
\begin{align}\label{eq:structfact2}
\gamma({\bf k}) = \frac12\,\left[ \cos (k_1) + \cos (k_2) \right ] .
\end{align}
Note that above we used the two-sublattice BZ of the {\it same} range
for both the honeycomb and square lattice---namely $-\pi<k_{1,2}\le\pi$.
The use of the same BZ enables us to compare easily any momentum-dependent function on the honeycomb and on the square lattice. Note that the structure factor for the honeycomb lattice can be obtained from the `bands' defined by Eq.~\eqref{eq:structfact1}
after substituting $k_1\rightarrow k_1/\sqrt{3}$ and $k_2\rightarrow 2 k_2/3$.

The important observation is that the modulus
of any of the four contributions to the quantum corrections to the order parameter $\Delta m$ \eqref{eq:deltamlswt}, i.e., $|C^\pm_{1, -1}|$, is
always by a factor $q\simeq 1.3$ times larger for the honeycomb lattice
than for the square lattice---i.e., just as $\Delta m$. Hence, despite
the fact that $C^\pm_{1, -1}$ have different signs and overall scales,
in order to understand why $\Delta m$
is $q$ times larger for the honeycomb than for the square lattice,
it is enough to investigate why the moduli $|C^\pm_{1, -1}|$ are always
$q$ times larger on the honeycomb lattice.

To this end, we plot in Figs.~\ref{fig:1}(a)-\ref{fig:1}(b)
and \ref{fig:1}(c)-\ref{fig:1}(d) the functions $C^\pm_{1, -1} ({\bf k})$
in the first BZ of the square and honeycomb lattice (note that the BZs are the same due to the rescaled momenta of the `standard' rectangular BZ of the honeycomb lattice, see above). We observe that for all momenta in the
substantial (central) part of the Brillouin zone the functions
$|C^+_{1}({\bf k})|$, $|C^-_{-1}({\bf k})|$, and $|C^+_{-1}({\bf k})|$
take all a higher value for the honeycomb lattice than for the square lattice.
It is only for relatively small areas around the corners of the BZ
[i.e., close to $(\pm \pi, \pm \pi)$ and $(\pm \pi, \mp \pi)$ momenta]
that the opposite situation takes place. At first sight a bit more intricate situation takes place for the $|C^-_{-1}({\bf k})|$ function, see Fig.~\ref{fig:1}(e).
In that case, one should also take into account the distinct singularities for both lattices. However, their contributions are in the end roughly equal
and we end up with a similar conclusion for $|C^-_{-1} ({\bf k})|$,
as for the case of \mbox{$|C^+_{1}({\bf k})|$, $|C^-_{-1}({\bf k})|$ and
$|C^+_{-1} ({\bf k})|$.}

In order to fully track the origin of the larger quantum corrections to the order parameter $\Delta m$
on the honeycomb lattice, we try to understand why in the large part of the BZ all of the `relevant' functions $|C^\pm_{1,-1}({\bf k})|$
take a higher value for the honeycomb lattice than for the square lattice.
To this end, we turn our attention to the moduli of the structure factors
$\pm |\gamma({\bf k})|$ for the honeycomb and square lattice---since these
are the crucial `quantities' entering Eqs.~(\ref{eq:c1})-(\ref{eq:c-1}).
Panels (c) and (f) of Fig.~\ref{fig:1} show
that in the large part of the BZ also the modulus of the structure factors
takes a higher value for the honeycomb lattice than for the square lattice.
Mathematically, this situation is to a large extent controlled by the fact that
the sets of zeros of the structure factor functions are very distinct for the honeycomb and for the square lattice: Whereas for the honeycomb lattice there are just two independent momenta for which $|\gamma({\bf k})|\equiv 0$ (at the
Dirac points), while for the square lattice there is a whole range of momenta for which $|\gamma({\bf k})|\equiv 0$
(note that the nesting property occurs along the magnetic BZ boundary).
Physically, this indicates the noticeably larger mobility of the spin waves on the honeycomb than on the square lattice.

\section{Summary and outlook}
\label{sec:summa}

Motivated by the recent studies of the van der Waals materials with
{\it quasi}-2D honeycomb magnetism we have investigated the onset of the
long-range order at $T=0$ of a continuous spin model on the honeycomb
lattice. Thereby we have shown that the order in the XY model occurs on
the honeycomb lattice but is there softer than for the square lattice:
Using the reflection positivity (RP) method we have shown that the
magnetic long-range order occurs for the honeycomb lattice in the XY
model for sufficiently large spin value $S\ge 2$. This stays in contrast
with the result obtained using the {\it same} method for the square
lattice which gives long-range order for spin $S\ge 1$~\cite{Kubo}.

The intuitive understanding of the above result can be achieved using
the (approximate) linear spin wave theory. We show that the enhanced
quantum spin fluctuations on the honeycomb lattice, as calculated using
the spin-waves are due to the overall much higher kinetic energy of the
spin waves on the honeycomb lattice than on the square lattice. The
latter can largely be traced back to the large qualitative differences
between the (nearest neighbor) structure factors: whereas on the
honeycomb lattice the structure factor has Dirac points and hence it
rarely vanishes, the good nesting properties of the square lattice
yield a whole range of momenta for which the structure factor vanishes.
We note that the obtained-here intuition goes beyond the simple argument,
which suggests that the order is less stable on the honeycomb than on the
square lattice due to the lower coordination number $z$ of the former
lattice (see Introduction).

We conclude by suggesting two open problems:
First, finding a rigorous proof of the existence of long range order
in the XY model on the honeycomb lattice {\it below} the spin value
$S=2$ remains a challenging open problem in the theory of magnetism.
It is a bit surprising that long range order has here this
constraint while a qualitative argument that this should be the
case is missing.

Second, rigorously verifying the existence of ordered state for the
Heisenberg model on the honeycomb lattice {\it with} small Kitaev
interactions may be an important, but supposedly also quite challenging,
exercise. So far, it is known that the interplay of Heisenberg and
Kitaev interactions gives an interesting phase diagram with several
ordered phases competing with spin liquids \cite{Cha13,Win17}, and an
experimental realization of the spin liquid was recently proposed
\cite{Tak19}. However, the border lines between particular phases depend
on the accuracy with which one treats quantum fluctuations
\cite{Got17,Mor18}. It would be interesting to control quantum
fluctuations in perturbation theory for the Kitaev-Heisenberg
(or Kitaev-XY) model with weak Kitaev interactions.

\begin{acknowledgments}
We thank Wojciech Brzezicki and Hosho Katsura for their interest and
insightful discussions. We kindly acknowledge financial support by
National Science Centre (NCN, Poland) under Project No.
2016/23/B/ST3/00839 (J. W. and A. M. O.) and
Project No. 2016/22/E/ST3/00560 (K. W.).
A. M. O. is grateful for support via the Alexander von Humboldt
Foundation \mbox{Fellowship} \mbox{(Humboldt-Forschungspreis).}

For the purpose of Open Access,
the authors have applied a CC-BY public copyright licence to any
Author Accepted Manuscript (AAM) version arising from this submission.
\end{acknowledgments}


%


\end{document}